\documentclass[12pt]{iopart}

\usepackage{graphicx}
\usepackage{float}
\usepackage{subfigure}
\graphicspath{{figures/}}

\bibstyle{unsrt}

\begin{document}

\title{Scattering model description of cascaded cavity configurations}
\footnote{This work was supported by the EU FP7 (ITN, CCQED-264666), the Hungarian National Office for Research and Technology under the contract ERC\_HU\_09 OPTOMECH, and the Hungarian Academy of Sciences  (Lend\"ulet Program, LP2011-016).}
\author{A.~Dombi and P.~Domokos}
\address{Wigner Research Centre for Physics, Hungarian Academy of Sciences, H-1525 Budapest, P.O.~Box 49}
\eads{\mailto{dombi.andras@wigner.mta.hu}, \mailto{peter.domokos@wigner.mta.hu}}

\begin{abstract}
Cascaded optical cavities appear in various quantum information processing schemes in which atomic qubits are sitting in separate cavities interconnected by photons as flying qubits. The usual theoretical description relies on a coupled-mode Hamiltonian approach. Here we investigate the system of cascaded cavities without modal decomposition by using a scattering model approach and determine the validity regime of the coupled-mode models.
\end{abstract}
\pacs{42.50.Pq, 42.50.Ex, 42.60.Da}

\noindent{\it Keywords\/}: coupled cavities, scattering model, quantum information processing

\maketitle

\section{Introduction}
Complex quantum information communication architectures can be composed of elementary blocks of cavity QED systems \cite{Wilk2007SingleAtom} wherein atoms are strongly coupled to the light field of a high-finesse resonator \cite{Cirac1997Quantum}. The q-bit can be stored in metastable internal states of a single atom, and can be efficiently mapped into the quantum state of a single photon inside the cavity \cite{Hijlkema2007Singlephoton,Bina2010Tripartite}. The photon can then be outcoupled, usually coupled into an optical fiber \cite{Trupke2007Atom}, and used as a flying q-bit to interconnect separate nodes. This principle has been demonstrated recently \cite{Ritter2012Elementary}. 

A very popular model describing the system of fiber connected cavities relies on the simplification that a single bosonic field mode represents the light field in the fiber (so called `short fiber limit') \cite{Serafini2006Distributed}, and this mode is linearly coupled to those of the cavities (coupled-oscillator model) \cite{Busch2010Generating}. One might suspect that when the cavity-fiber interfaces redefine the radiation field in the fiber such as it can be described in terms of discrete longitudinal modes (instead of the normal continuum), then all the boundary conditions within the setup together define the mode structure. That is, in the simplest case, two coupled, linear cavities should be considered as an  interferometer consisting of four mirrors. In this paper we will investigate  such a four-mirror interferometer by means of a fundamental scattering approach \cite{Asboth2008Optomechanical,Xuereb2010Optomechanical}. We aim at checking the coupled-oscillator model:  determine its domain of validity and, eventually, uncover possible artifacts which might wrongly be built into a quantum information processing protocol.

\section{Cascaded cavities}
\begin{figure}[htb]
\begin{center}
	\includegraphics[width=0.5\textwidth]{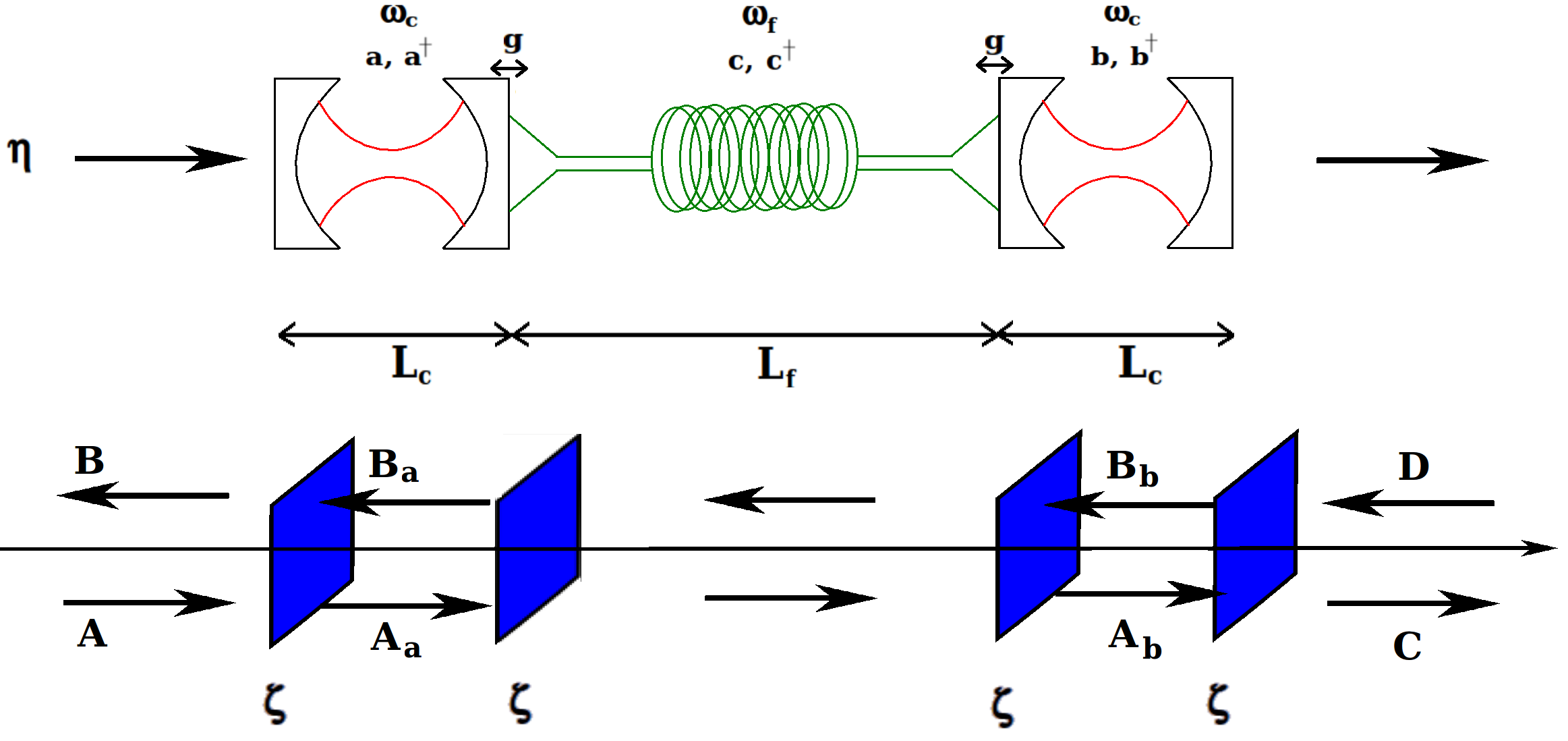}
	\caption{Above: Two identical cavities characterized by the resonance frequency $\omega_C$ and an effective coupling strength $g$ to a single mode of a fiber. Below: Four mirrors with separations $L_C$ on the sides, corresponding to the cavity length, and $L_F$ in the middle, corresponding to the fiber length. The mirrors are identical and can be characterized by the complex polarizability parameter $\zeta$.}
	\label{fig:CascadedCavitiesScheme}
\end{center}
	\end{figure}

The system we are studying is schematically depicted in Figure \ref{fig:CascadedCavitiesScheme}. It consists of two identical cavities (optical Fabry-Perot resonators), which are coupled via an optical fiber. The same system can be viewed as an ensemble of four mirrors. Considering one-dimensional propagation only, the fact that the cavities are connected by fiber does not play an essential role. 

\subsection{Coupled oscillator model}

In the case of the coupled oscillator model we consider single-mode fields both in the cavities and in the fiber, the corresponding bosonic annihilation operators are denoted by $a$, $b$ and that of the fiber mode by $c$. The resonant frequency of the cavities is $\omega_C$ and that of the fiber is  $\omega_F$. The Hamiltonian of the system,  in units of $\hbar=1$, is  
	\begin{equation}
	H =\omega_C a^\dagger a + \omega_C b^\dagger b + \omega_F c^\dagger c  + \, g \,  (a^\dagger c+c^\dagger a) + \, g \, (b^\dagger c+c^\dagger b) \; ,
	\end{equation}
where the linear coupling between the fiber and the cavities is described by the effective coupling constant $g$. The system can be driven from both directions, with effective pump amplitudes $\eta_L$ and $\eta_R$, respectively. The laser driving frequency is $\omega$, so the pump Hamiltonian is
	\begin{equation}
	 \label{eq:H_pump}
	H_{\rm pump} = \eta_L \,  \left(a^\dagger e^{-i \omega t}+ a e^{i \omega t}\right) + \, \eta_R \,  \left(b^\dagger e^{-i \omega t} \, e^{-i \varphi} + b e^{i \omega t} \, e^{i \varphi}\right)\; ,
	\end{equation}
where $\varphi$ accounts for a phase difference between the left and right pump amplitudes. We assume perfect mirrors so that the only photon loss source is transmission into the free space.  This loss is described by the following dissipative terms in the quantum master equation
	\begin{equation}
	\dot{\rho} = i\left[\rho, H\right]  - \kappa \left(a^\dagger a \rho + \rho a^\dagger a - 2 a \rho a^\dagger\right)\;,
	\end{equation}
and similar terms for the cavity mode $b$.  The cavity photon loss rate is $2 \kappa$. Note that the loss of an isolated Fabry-Perot cavity occurs through both mirrors.

Such a simplified model is needed when the dynamics of a more complex system including  atomic q-bits is considered \cite{Ogden2008Dynamics,Montenegro2012Entanglement,Alexanian2011Twophoton,ShiBiao2011Entangling}. Then additional atom-field interaction terms have to be included, of course. However, it remains a question if such a simplified treatment of  the field itself is justified.

\subsection{One-dimensional scattering model}

The four mirror interferometer is analyzed by means of the one-dimensional scattering model  and the transfer matrix method \cite{Asboth2008Optomechanical,Xuereb2010Optomechanical}. In every point along the optical axis, the field is described by the left- and right propagating plane wave amplitudes. In the transfer matrix method these amplitudes on the left and right sides of an optical element (`scatterer') are related to each other by a linear matrix:  
\begin{equation}
	\left(\begin{array}{c}						
		 C\\
		 D						
		\end{array} \right)= M  \left(\begin{array}{c}
		 A\\
 		 B
		\end{array} \right)\; ,
\end{equation}
where $C$, $D$ are on the right and $A$, $B$ are on the left side of a scatterer; $A$, $C$ are the right and $B$, $D$ are the left propagating plane wave mode amplitudes.
There are two types of matrices we need to describe the four-mirror interferometer. First, that of a mirror,
\begin{equation}
	M_{\rm mirror} = \left[ \begin{array}{cc}
		1-i\zeta & -i\zeta \\
		i\zeta & 1+i\zeta \end{array} \right] \; ,
\end{equation}
where the mirror is characterized by the single parameter of linear polarizability $\zeta$, which defines the reflectivity $r = \frac{i\zeta}{1-i\zeta}$ and  transmissivity $t = \frac{1}{1-i\zeta}$. Here again, we will consider absorption free, perfect beam splitters as mirrors, which amounts to having real polarizability parameter $\zeta$. Second, the free propagation between the mirrors, which is given by
\begin{equation}
	M_{\rm prop} = \left[ \begin{array}{cc}
	e^{ikd} & 0\\
	0 & e^{-ikd} \end{array} \right]\; ,
\end{equation}
where $k$ is the wavenumber of the plane wave of frequency $\omega$, and $d$ is the distance between the two mirrors.

\subsection{Matching the parameters}

In the following we will match the parameters on the basis of deriving the same, well-defined, measurable physical quantity in both of the models. We will make use of the fact that the correspondence between the parameters must be independent of the geometry. Therefore, we can consider a simplified setup to establish the connection. 

\subsubsection{{Single cavity transmitted photo-current}}

First, let us look at the transmitted photo-current of a single, driven cavity. The driven and lossy oscillator model gives for the outgoing photon number per unit time 
\begin{equation}
  j_{\rm out} =  \kappa \langle a^\dagger a \rangle = \frac{\kappa |\eta|^2}{(\omega-\omega_C)^2 + \kappa^2}\; ,
\end{equation}
which is the simple Lorentzian spectrum of a resonator. Remember that only half of the outgoing photons leaves into one given direction (to the right). The transmitted photon number in the scattering model is 
\begin{equation}
  \label{eq:Photocurrent}
 j_{\rm out} = \frac{2S\epsilon_0 c}{\hbar \omega} |C|^2 \; ,
\end{equation}
where $S$ is a (usually fictitious) quantization surface, perpendicular to the optical axis. The outward propagating field amplitude$C$ can be obtained by straightforward calculation. Close to resonance, the spectrum can be approximated by a Lorentzian, which gives rise to the relations
\begin{eqnarray}
 \kappa &=& \frac{c}{L_C} \, \frac{1}{2 \zeta \sqrt{\zeta^2+1}}\\
 \omega_C &=& \frac{c}{L_C} \left[ n \, \pi + \frac{1}{2} {\rm atan}\left\{ \frac{2 \zeta}{1-\zeta^2}\right\} \right] \\
 \eta_L &=& \sqrt{\kappa} A \; ,
\end{eqnarray}
where the scattering model parameters $L_C$ (cavity length) and $\zeta$ are used, and $n$ is an integer number giving the order of the resonance. The pumping amplitude from the left side, $\eta_L$, is of course related to the incoming field amplitude $A$ of the scattering model.

\subsubsection{Coupled cavity resonances}

We still need to determine the coupling constant $g$ of the coupled-cavities model. To this end, we consider the resonances of a system of two coupled cavities which have different lengths, $L_1$ and $L_2$, but have a common resonance frequency $\omega_C$. 
Using the same notations as before, the Hamiltonian of the system is $ H=\omega_C a^\dagger a + \omega_C b^\dagger b + g (a^\dagger b + b^\dagger a) $, and the eigenfrequencies are 
 $\omega_\pm= \omega_C \pm g  \,.$
That is, the splitting is $2g$. For the three-mirror system, lengthy but straightforward calculation within the scattering model leads to the normalized transmission spectrum  
\begin{equation}
\frac{\left|C  \right|^2}{\left| A \right|^2} = \left[1 + \zeta^2 (1+\zeta^2)^2 \left( 4 \cos^2\left\{\frac{kL_1 +kL_2}{2}- \Phi/2 \right\} -\frac{1}{1+\zeta^2}\right)\right]^{-1}\; ,
\end{equation}
where $\Phi$ is a global shift of the spectrum. We find that the splitting between resonances can be expressed by the linear polarizability $\zeta$, the lengths of the cavities $L_1$ and $L_2$ as 
\begin{equation}
 2 g= \frac{2 c}{ (L_1+L_2) \sqrt{1+\zeta^2}}\; .
\end{equation}
This expression can be applied to the case of the coupling of the fiber and a cavity.

\section{Comparison of the transmission spectrums}

\begin{figure}[H]
\centering
\mbox{\subfigure[]{\includegraphics[width=0.5\textwidth]{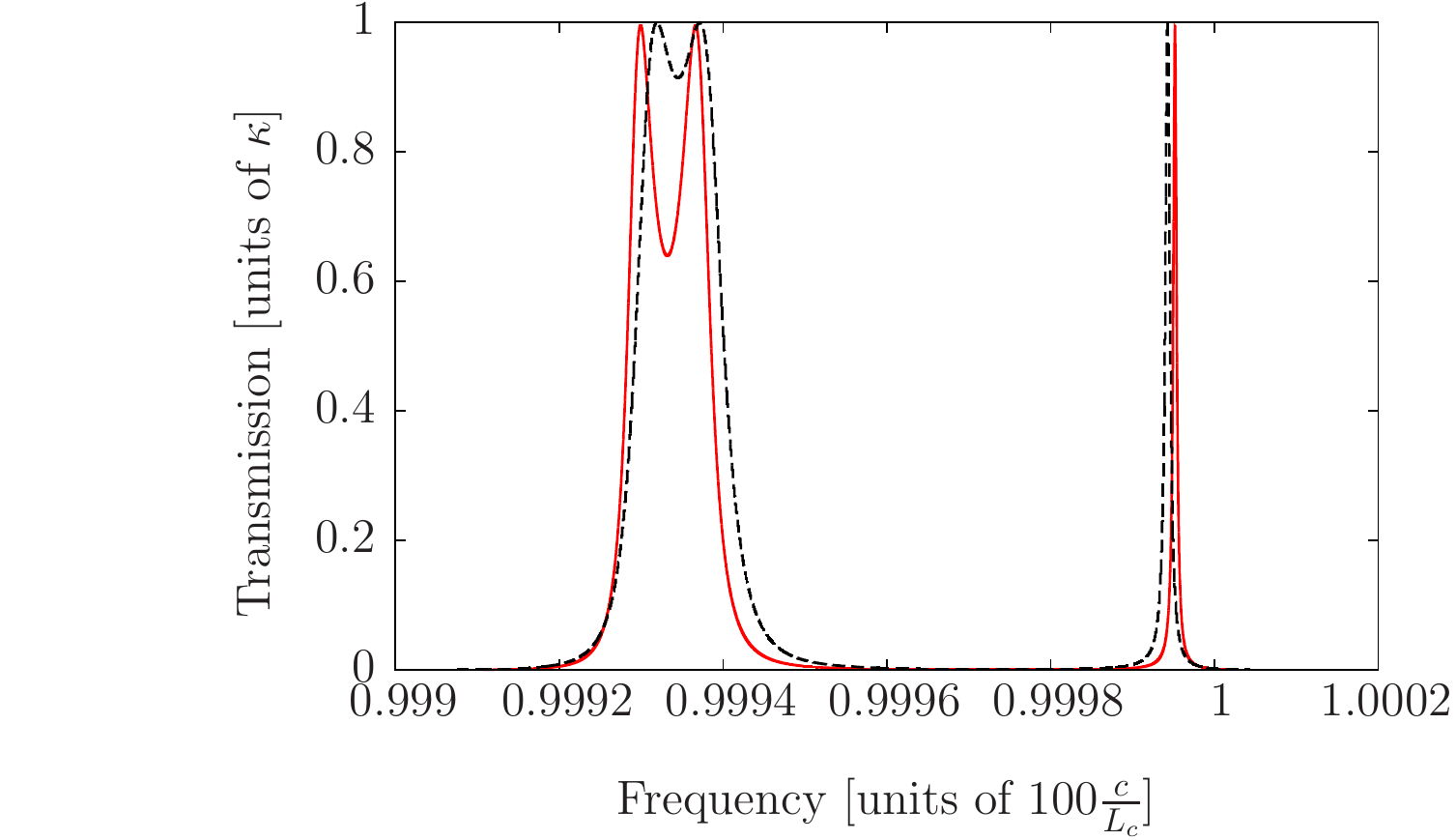}}\quad
\subfigure[]{\includegraphics[width=0.54\textwidth]{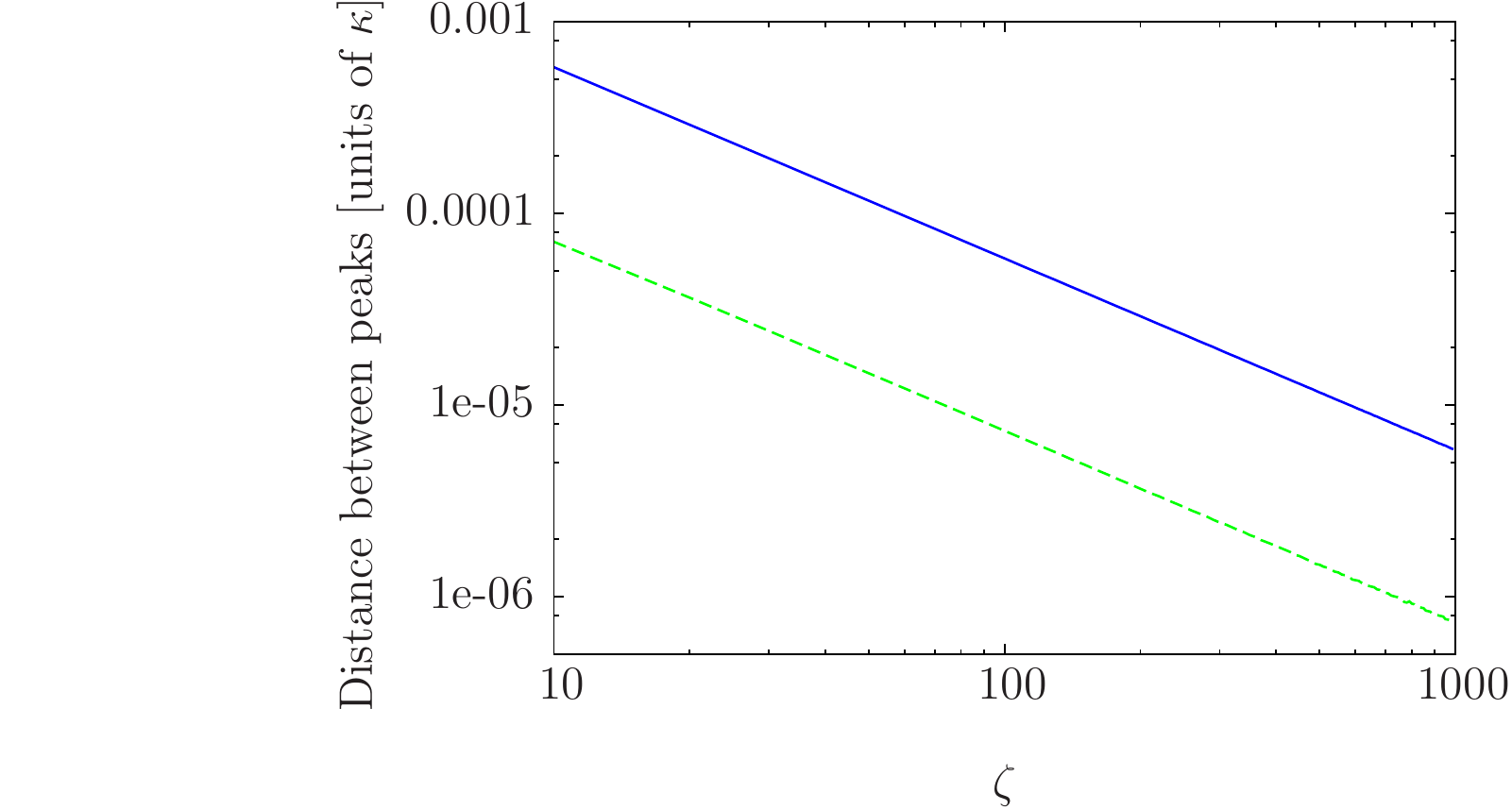} }}
\caption{(a) Transmitted photocurrent through the two cascaded cavities according to the coupled oscillator model (red) and scattering model (black). The cavities are separated by a distance of $L_F=5L_C$ where $L_C$ is the cavity length. The mirrors are characterized by $\zeta=5$ which corresponds to the reflectivity $|r|^2 \approx 0.96$. The three peaks correspond to three resonances of the system around the bare cavity frequency $\omega_C$. Other peak families (not shown) exist at integer multiples of the free spectral range away.  (b) The distances of the side peaks from the middle one are not the same in the two models, their difference is plotted here.} 
\label{fig:TransmissionSpectra}
\end{figure}

After matching the parameters, we can return to the problem of cascaded cavities and quantitatively compare the results of the two models.  In the first step we look at the transmission spectrum which shows the positions and widths of the resonances. In Fig.~\ref{fig:TransmissionSpectra}a, we plot the transmitted photocurrent as a function of the pump laser frequency $\omega$. The mean photocurrent is $\kappa \langle b^\dagger b \rangle$ in the coupled oscillator model, whereas in the scattering model it is given by Eq.~(\ref{eq:Photocurrent}), and the amplitude $C$ has to be calculated according to the geometry by the transfer matrix method.
The difference between the two models is explicitly represented in Fig.~\ref{fig:TransmissionSpectra}b, where the separation of the resonance peaks are plotted as a function of mirror polarizability. This representation  reveals that the coupled-oscillator model gets better and better with increasing  the linear polarizability  $\zeta$, i.e., in the good cavity limit. In the plotted range, the transmissivity of the mirrors is below 1\%, and the spectra overlap (apart from some possible uninteresting offset) within a small fraction of the cavity line width $\kappa$.

\section{Spatial distribution of the field in the cascaded cavity setup}

In a quantum communication setup the q-bits couple to the field in their position and the local field intensity is the essential quantity which determines the atom-field coupling.  In the oscillator model the modes $a$, $b$ and $c$ are confined to the spatial domains of the respective cavities and the fiber. The eigenmodes of the total system, however, are spatially delocalized, and the resonant excitations couple simultaneously to different qubits. Therefore, it is important to analyze the spatial distribution of the field associated with the resonances. 

\begin{figure}[H]
\centering
\mbox{\subfigure[]{\includegraphics[width=0.5\textwidth]{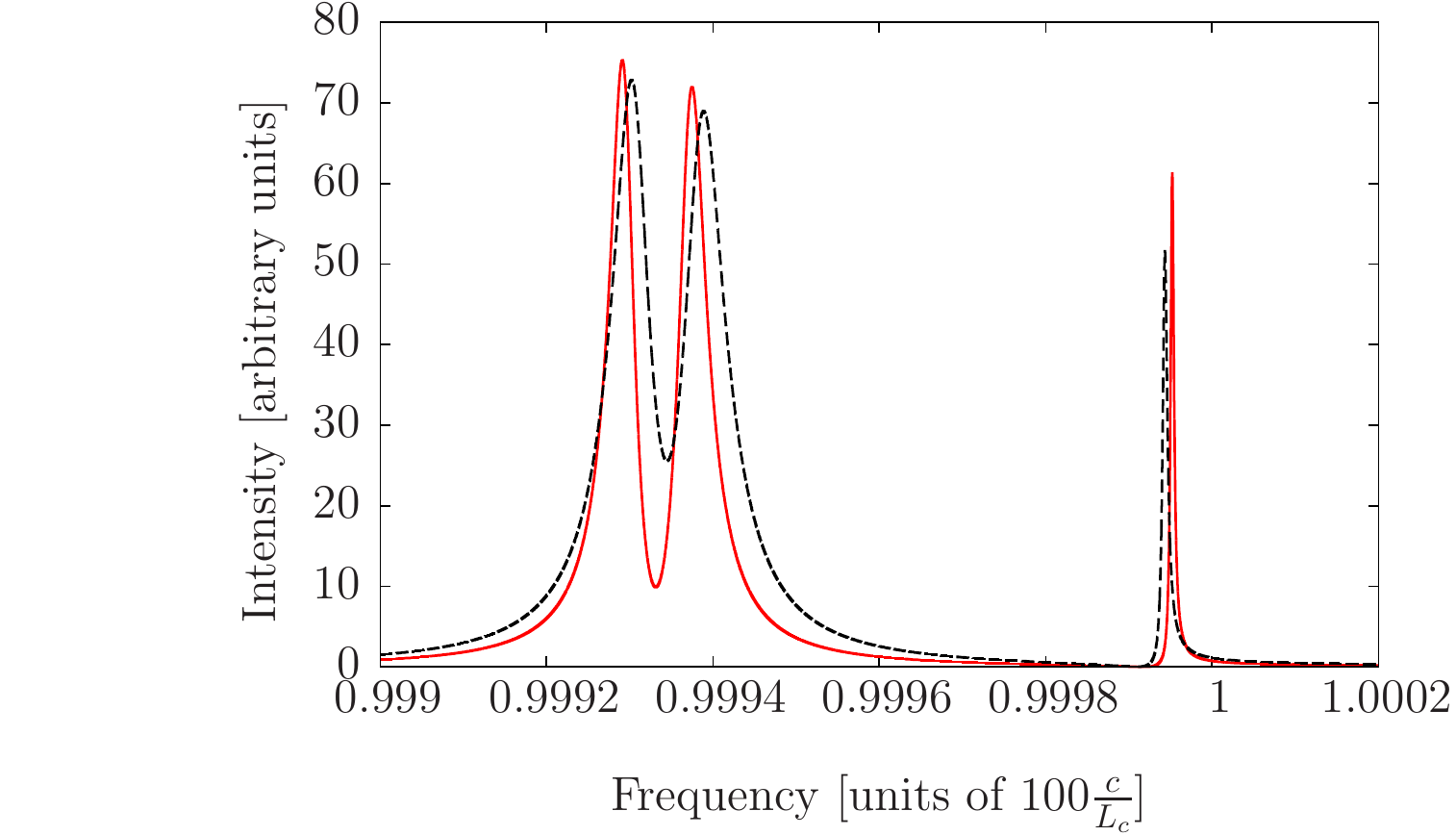}}\quad
\subfigure[]{\includegraphics[width=0.5\textwidth]{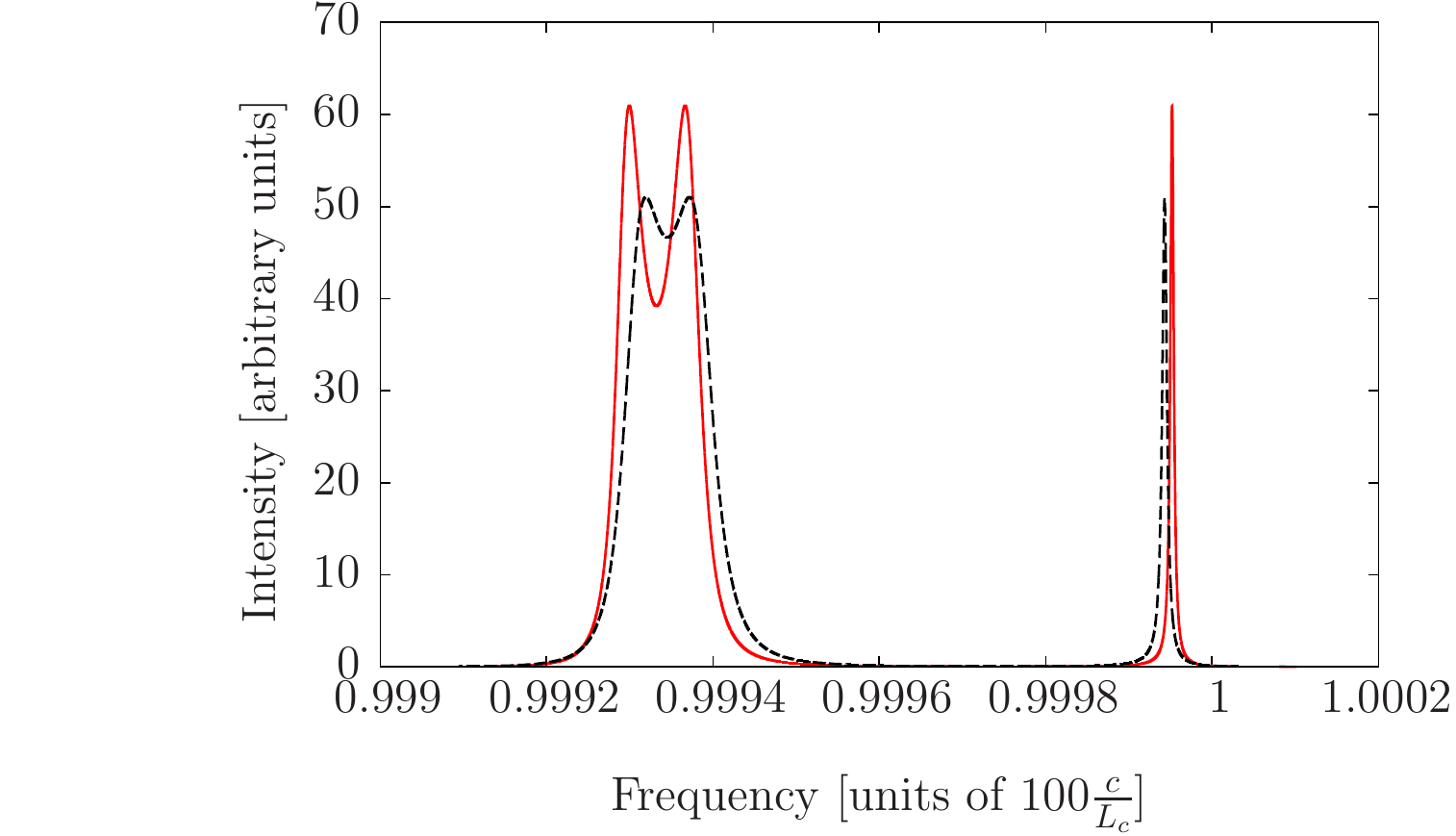} }}
\caption{Intensities (a) in the left and (b) in the right cavities for the cascaded cavity driven from the left, calculated from  the coupled oscillator (solid red lines, $\langle a^\dagger a\rangle$ and $\langle b^\dagger b\rangle$) and the scattering models (solid black lines, $|A_a|^2 + |B_a|^2$ and $|A_b|^2 + |B_b|^2$). Parameters are the same as for the transmission spectra in Fig.~\ref{fig:TransmissionSpectra}(a): $\zeta=5$, $L_F=5L_C$.} \label{fig:SpatialDistrib}
\end{figure}
As a first example, in Fig.~\ref{fig:SpatialDistrib}(a) and (b) we plot the cavity photon number in the left and right cavities, respectively, corresponding to the transmission spectrum of Fig.~\ref{fig:TransmissionSpectra}. Obviously, the plot of the local field intensities reflects the same difference in the position of the resonant frequencies as found in the transmission spectrum. In addition, we get a difference in the magnitude of the intensities in the two models. This means that the true modes (associated with the observable resonances) have a spatial distribution that cannot be mimicked by the coupled oscillator model, so it does not provide for the precise values of the coupling to a qubit.

\subsection{Are there dark modes?}

\begin{figure}[H]
\centering
\mbox{\subfigure[]{\includegraphics[width=0.5\textwidth]{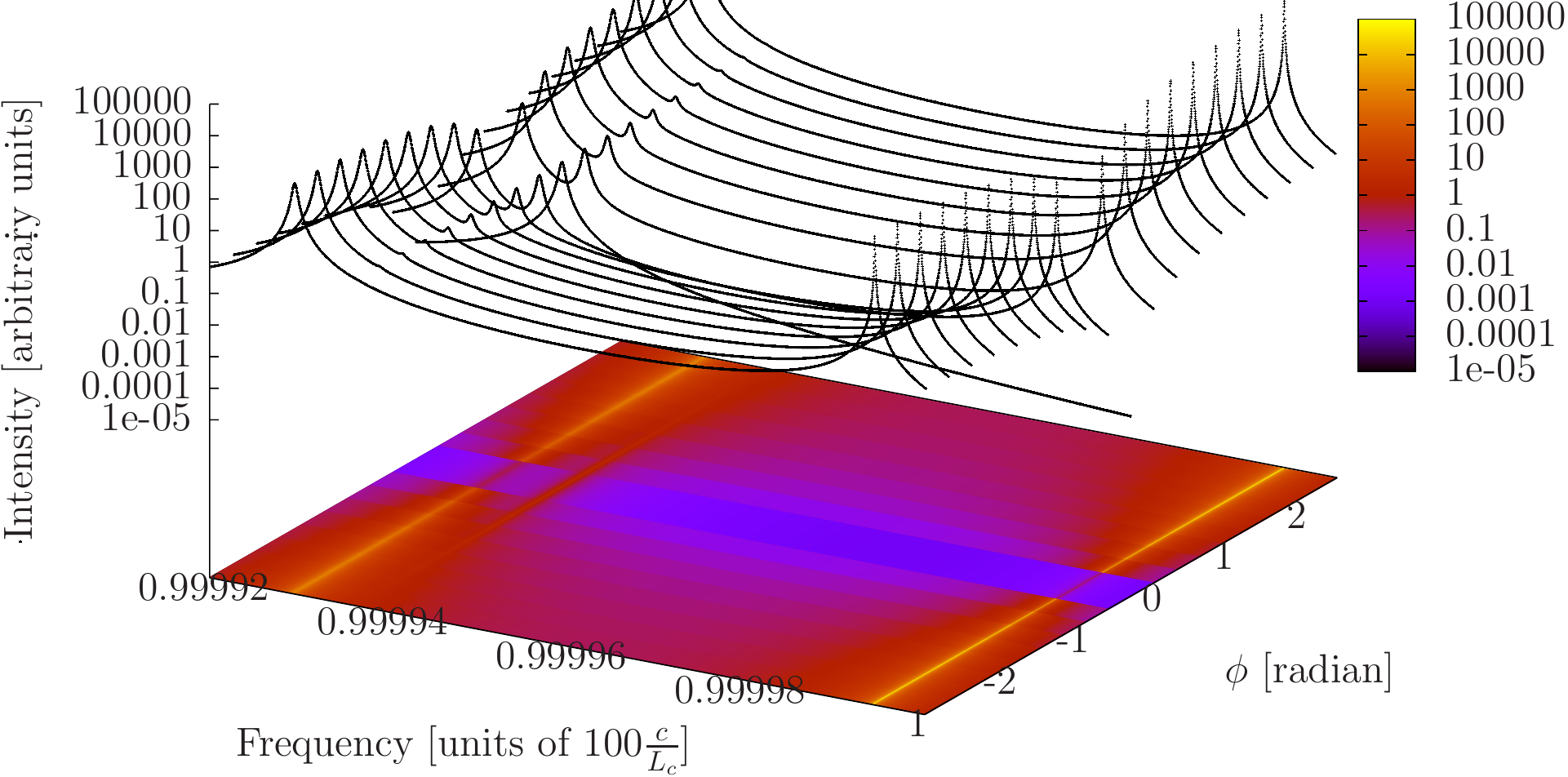}}\quad
\subfigure[]{\includegraphics[width=0.5\textwidth]{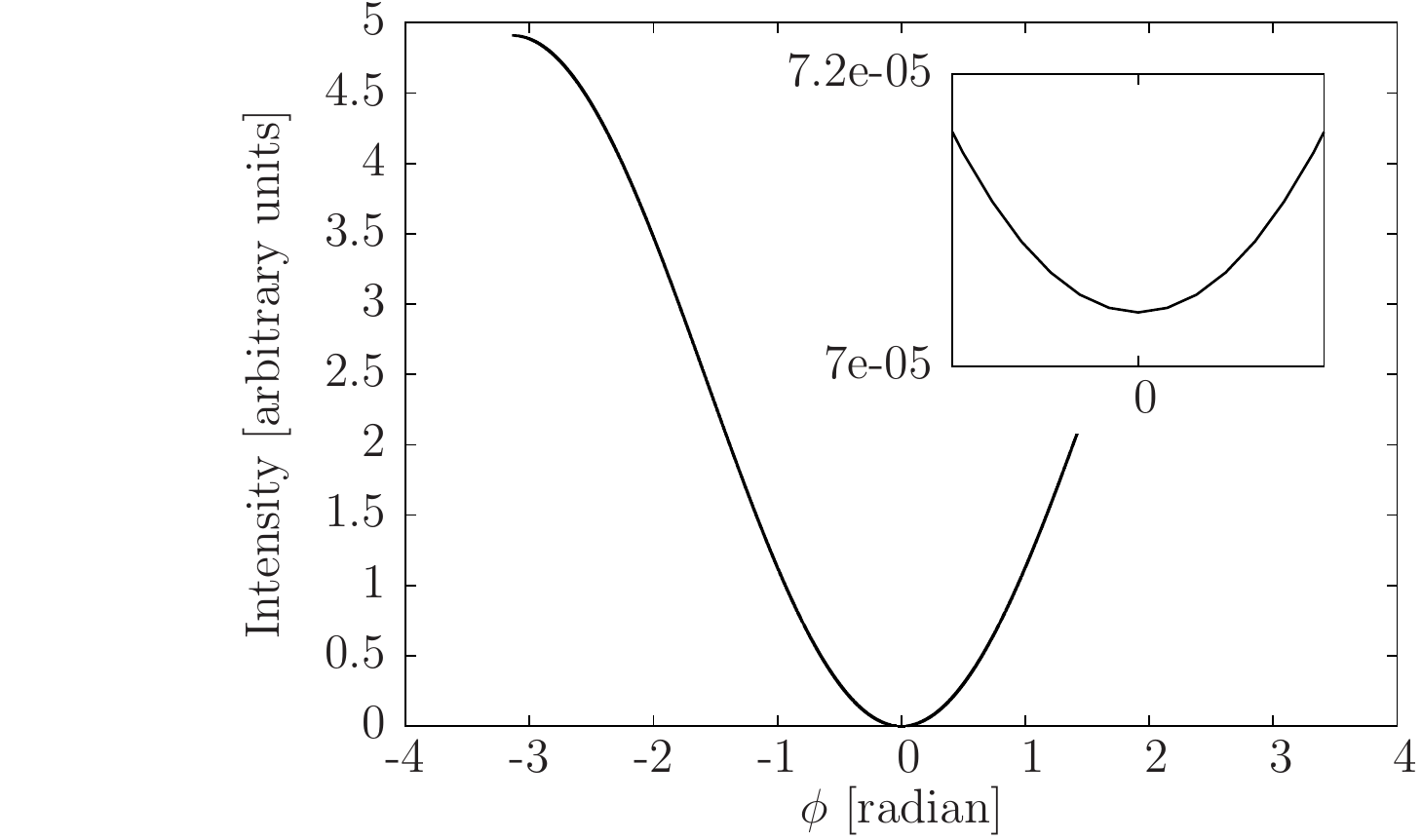} }}
\caption{(a) The intensity (logarithmic scale) in the fiber for incoming amplitudes $A=1$ and $D=e^{ -i\phi}$  from the left and right sides as a function of the pump frequency and the phase $\phi \in [-\pi, +\pi]$. Parameters: $\zeta=5$, $L_F=5L_C$. (b) Cut of the surface plot at the pump frequency 0.9999 in order to show the sinusoidal phase dependence and the almost perfect destructive interference (in the inset).} \label{fig:PumpPhase}
\end{figure}

There is an interesting possibility to demonstrate the difference between the spatial distributions obtained form the two models. The coupled oscillator model allows for the occurrence of dark modes for certain linear combinations of a two-sided pumping. In particular, if the left and right cavities are pumped with the same amplitude but opposite phase ($\eta_L=\eta_R$, $\varphi=\pi$ in Eq.~(\ref{eq:H_pump})), the symmetric mode $a+b$ decouples from the pump. Furthermore, this symmetric mode couples to the fiber mode $c$, therefore the field in the intermediate domain should vanish. In principle exact suppression is expected. Note that this is impossible in the scattering model: if both propagating field mode amplitudes are exactly zero at any point, the field must vanish everywhere. Figure \ref{fig:PumpPhase}(a) shows the intensity in the fiber as a function of the pump frequency (spectrum) and also as a function of the relative phase between the left and right pumps. 
The fiber intensity is resonantly enhanced for the two side resonances, in accordance with the coupled oscillator model. The third, much smaller resonance in between is absent in the coupled oscillator model in which this mode does not involve the mode $c$. In the whole frequency range,  the intensity depends sinusoidally on the relative phase $\phi$. As can be seen in \ref{fig:PumpPhase}(b), the destructive interference is almost perfect at $\phi=0$.

In conclusion, we showed that the cascaded cavities configuration can be well treated by the very simple coupled oscillator model in the good cavity limit, e.g., provided the reflectivity of the mirrors is above 99\%. However, one must be careful when coupling of the local field to atomic qubits is considered: no local addressing is possible, the field extends in the whole setup with a highly non-trivial manner reflecting interferometric sensitivity to the parameters. This might lead to effects detrimental to a given protocol, if not taken into account a priori. 
\\
\bibliographystyle{unsrt}
\bibliography{cascadedcavity}

%
%
%
\end{document}